\documentclass{sf2a-conf2017}
\usepackage{graphicx}
\usepackage{hyperref}
\usepackage[]{natbib}  
\usepackage{epstopdf}

\def\BibTeX{{\rm B\kern-.05em{\sc i\kern-.025em b}\kern-.08em
    T\kern-.1667em\lower.7ex\hbox{E}\kern-.125emX}}
\bibpunct{(}{)}{;}{a}{}{,}  %%%%%%%%%%%%%  A&A bibliography style
\begin{document}

\TitreGlobal{SF2A 2017}

%%-----------------------------------------------------------------
%%      the top matter
%%

\title{The impact of Gaia on our understanding of \\ the Vast Polar Structure of the Milky Way}

\runningtitle{Gaia and the VPOS}

\author{Marcel S. Pawlowski}\address{Hubble Fellow, Department of Physics and Astronomy, University of California, Irvine, CA 92697, USA}

%% Keep this line, even if the page will be settled afterwards.
\setcounter{page}{237}

%%-----------------------------------------------------------------

\maketitle

%%-----------------------------------------------------------------
%%        The abstract
%% 
%%  Warning!  within the abstract:
%%  - do not use macros. 
%%  - do not use commands like: \cite, \citet, \citep ... etc.

\begin{abstract}
The Milky Way (MW) is surrounded by a Vast Polar Structure (VPOS) of satellite galaxies, star clusters, and streams. Proper motion measurements for the brightest MW satellites indicate that they are predominantly co-orbiting along the VPOS. This is consistent with a dynamically stable structure. Assuming that all satellites that are aligned with the VPOS also co-orbit along this structure allows to empirically predict their systemic proper motions. Testing predictions for individual satellite galaxies at large distances requires high-accuracy proper motion measurements such as with the Hubble Space Telescope. However, for nearby MW satellites, Gaia will allow to test these predictions, in particular for a statistical sample of satellites since proper motion predictions exist for almost all of them. This will clarify how rotationally supported the VPOS is. In addition, Gaia will discover Galactic substructure, in particular stellar streams. The degree of their alignment with the VPOS might further constrain its stability and nature. 
\end{abstract}

%% Insert the keywords (to appear in the ADS indexing)
%% Keywords must be separated by a comma

%%-----------------------------------------------------------------

\section{Introduction}
%%---------------------

\begin{figure}[ht!]
 \centering
 \includegraphics[width=0.8\textwidth,clip]{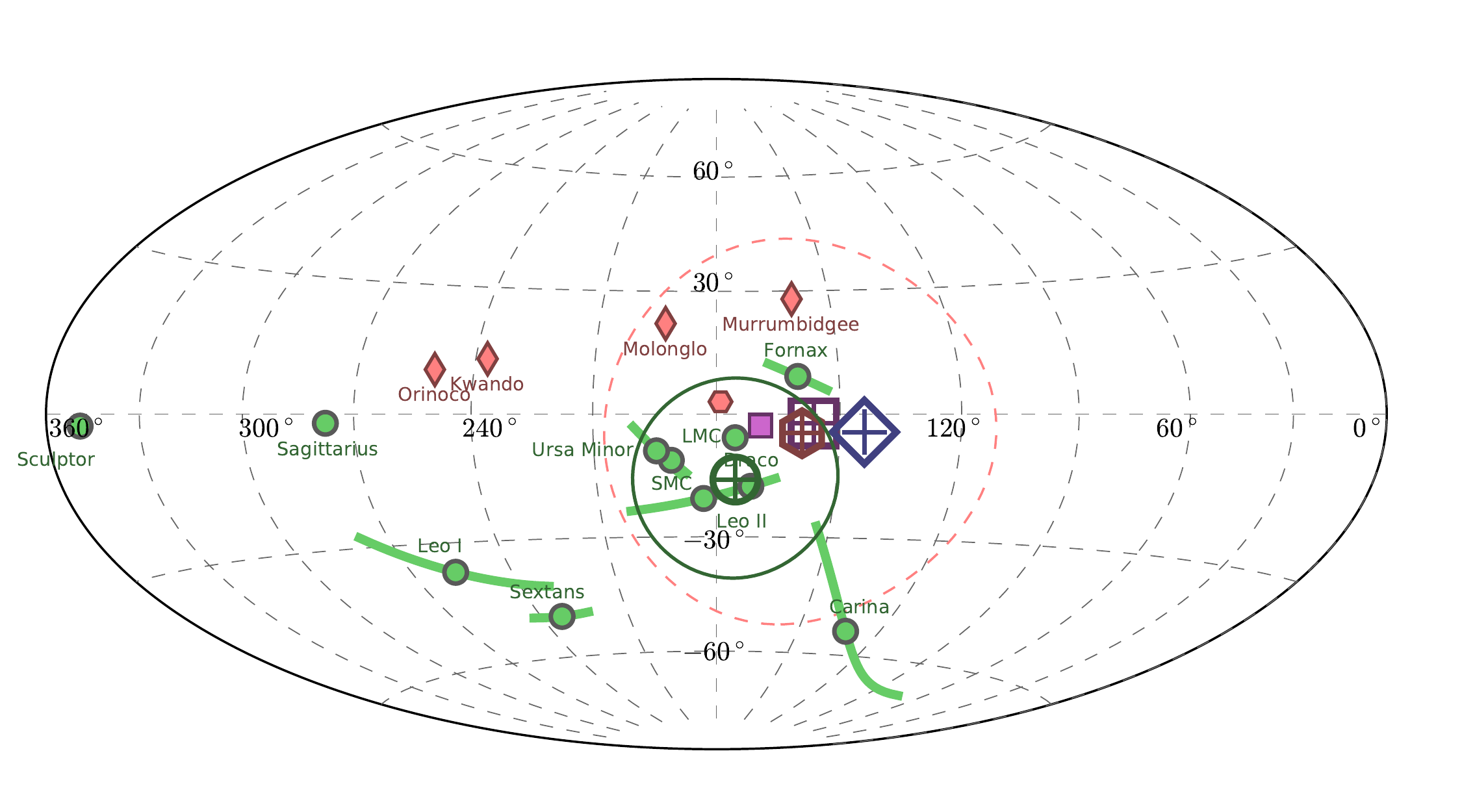}      
%% Note the ABSENCE of the extension .pdf  !
  \caption{All-sky plot showing the directions of the orbital poles for the 11 classical satellite galaxies for which proper motions have been measured (green dots with $1\sigma$\ uncertainties). The figure is based on \citet{Pawlowski2013}, and has been updated with new proper motion measurements, most recently by \citet{Sohn2017} and \citet{Casetti-Dinescu2017}. 
The normal directions to the planes fitted to the positions of satellite galaxies (magenta squares) and distant globular clusters thought to have been associated with accreted dwarf galaxies (blue diamond) lie close to the strong concentration of orbital poles, indicating that many of the satellite galaxies co-orbit in the VPOS. Also shown are the average of normal vectors to streams in the MW halo (red, from \citealt{Pawlowski2012} and updated in \citealt{PawlowskiKroupa2014}) and the Magellanic Stream (small read hexagon). The orbital poles of four stellar stream candidates recently found by \citet{Grillmair2017} in the Galactic south (red, with names) are consistent with the finding that about half of the distant streams in the MW halos are oriented approximately along the VPOS. 
  }
  \label{pawlowski2:fig1}
\end{figure}

The satellite galaxies of the Milky Way (MW) are distributed anisotropically. This was first noted by \citet{Kunkel1976} and \citet{Lynden-Bell1976}, who demonstrated that the then-known satellites align along a common great circle, which is also shared by the Magellanic Stream. Already then this lead to the speculation that at least some of the MW satellite galaxies move along as part of a coherent structure.
The subsequent discovery of additional satellite galaxies has corroborated this early finding (see \citealt{Pawlowski2015} for an analysis including the most recently discovered satellite galaxies). The satellite galaxies of the MW are preferentially aligned in a flattened distribution oriented perpendicular to the Galactic disk. This Vast Polar Structure (VPOS) also contains distant globular clusters, and about half of the known stellar and gaseous streams of satellite systems disrupting in the Galactic halo are aligned with it \citep{Pawlowski2012}. Since tidal debris streams trace the orbital plane of their progenitors, this indicates a preference for satellite systems to orbit along the VPOS, whose orientation is defined by the current satellite positions. Proper motion measurements, available for all of the 11 classical satellite galaxies, reveal that this is indeed the case \citep{Metz2008,Pawlowski2013}. Eight of the 11 satellites are consistent with orbiting in the VPOS, with Sculptor being the only one in a counter-orbiting orientation. 

This intriguing alignment is illustrated in Figure \ref{pawlowski2:fig1}, which plots the normal vectors defining the orientations of planes fitted to the positions of satellite galaxies, globular clusters, and streams, as well as the orbital poles (directions of orbital angular momentum) of individual satellite galaxies. It includes updated and new proper motion measurements for several satellite galaxies \citep[e.g.][]{Sohn2017,Casetti-Dinescu2017}, and also the orbital poles of four stellar stream candidates discovered by \citet{Grillmair2017}. Interestingly, the two streams of these which best align with the VPOS are those with the largest pericentric distances ($\geq 20$\,kpc): Molonglo and Murrumbidgee. The other two have pericentric distances below 7\,kpc and thus their orbits can be expected to show more precession.

% Since the existence of the VPOS and similar structures around other galaxies is in tension with expectations based on $\Lambda$CDM cosmology, it is even more important to determine the ???

\section{Predictions for systemic proper motions of Milky Way satellite galaxies}
%%-------------------------
\begin{figure}[ht!]
 \centering
 \includegraphics[width=0.54\textwidth,clip]{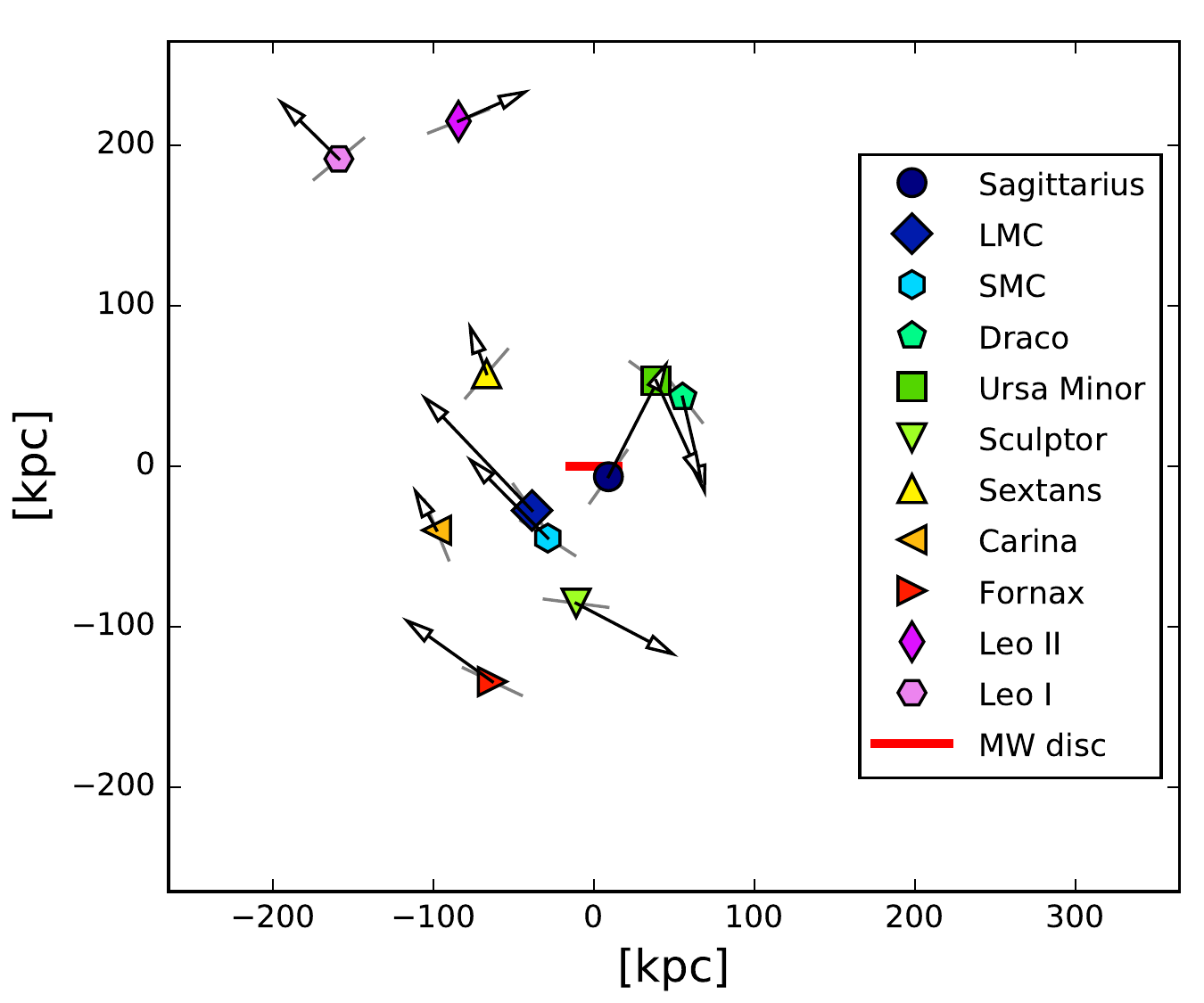}      
 \includegraphics[width=0.28\textwidth,clip]{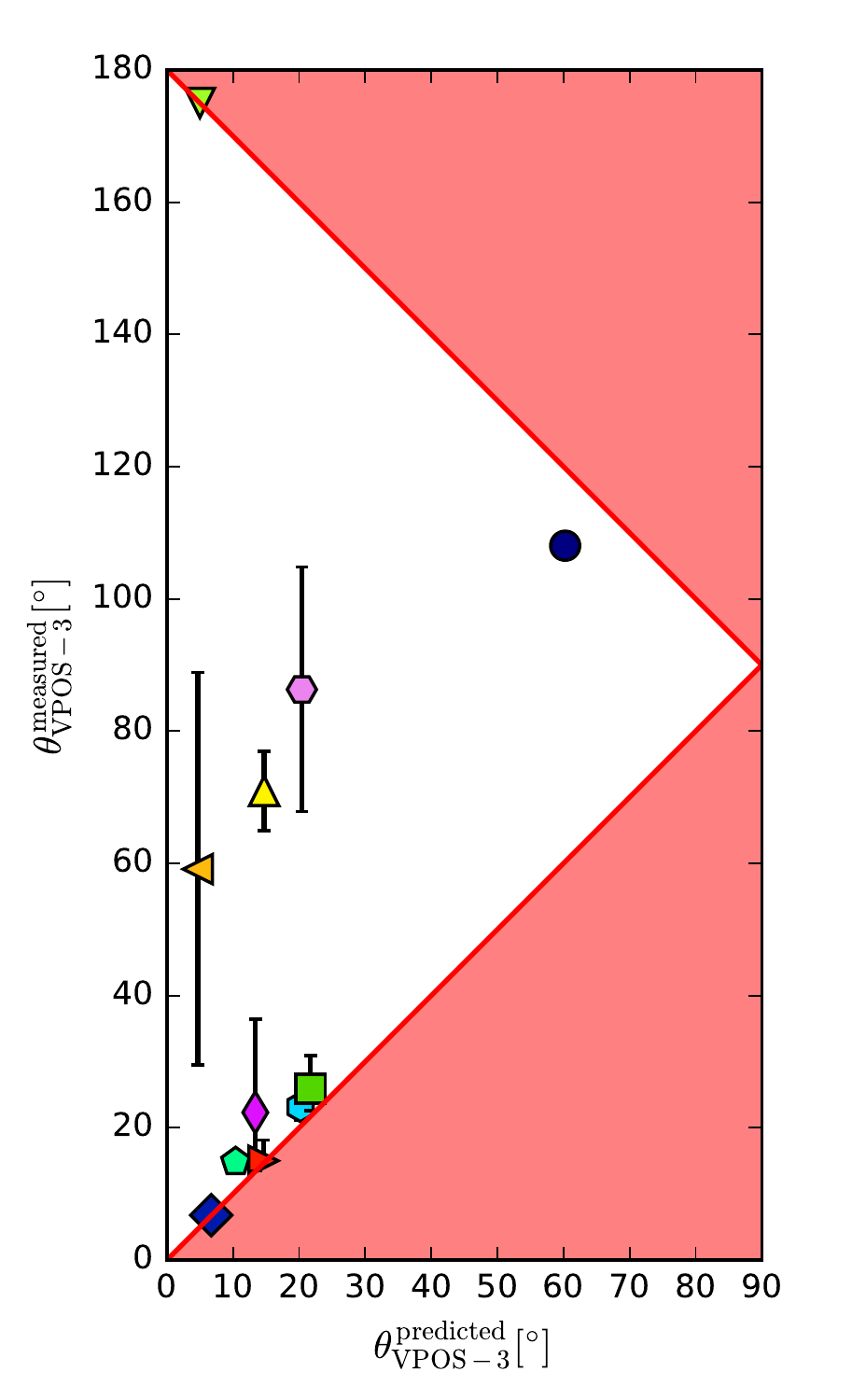}      
%% Note the ABSENCE of the extension .pdf  !
  \caption{{\bf Left:} The distribution of the 11 classical satellite galaxies around the MW, in a view oriented face-on relative to the VPOS. The arrows show their velocities, based on the measured line-of-sight velocities and proper motions. There is a clear preference for the satellites to co-orbit, as well as for an excess of the tangential over the radial velocity component. {\bf Right:} Predicted (minimum possible) angle between a satellite's current orbital plane and the VPOS $\theta_{\mathrm{VPOS-3}}^{\mathrm{predicted}}$, plotted against the corresponding angle $\theta_{\mathrm{VPOS-3}}^{\mathrm{measured}}$\ based on the measured 3D velocity of the satellite. Most satellites with well determined proper motions (small error bars) lie close to the prediction (red lines). Same symbols as in the left panel. Both plots are updated versions based on the analysis presented in \citet{Pawlowski2013}.}
  \label{pawlowski2:fig2}
\end{figure}

Since there is empirical evidence that the VPOS is, at least in part, rotationally stabilized, one can formulate the hypothesis that satellites which align with the structure also orbit along it. This in turn results in an empirical prediction of the proper motion that would make a given satellite orbit as closely along the VPOS as possible. A detailed description of how these predictions are made can be found in \citet{Pawlowski2013}. 

In brief, for a given satellite the angle between its position vector from the Galactic center and the orientation of the best-fit VPOS plane gives the best possible orbital alignment. The corresponding orbit requires that the 3D velocity of the satellite is confined to lie in this orbital plane. Since the position, distance, and line of line-of-sight velocity of the satellite is known, this results in a linear relation between the two proper motion components that constitutes the range of predicted proper motion. The maximum and minimum tangential velocities are further constrained, respectively, by requiring the satellite to be gravitationally bound to the MW (otherwise it would not be a satellite in the strict sense), and by requiring that the orbit is not predominantly radial (in which case the orbital orientation becomes meaningless). Predictions for a total of 36 MW satellite galaxies can be found in \citet{Pawlowski2013}, \citet{PawlowskiKroupa2014}, and \citet{Pawlowski2015}.

\section{Testing proper motions}

\begin{figure}[ht!]
 \centering
 \includegraphics[width=0.32\textwidth,clip]{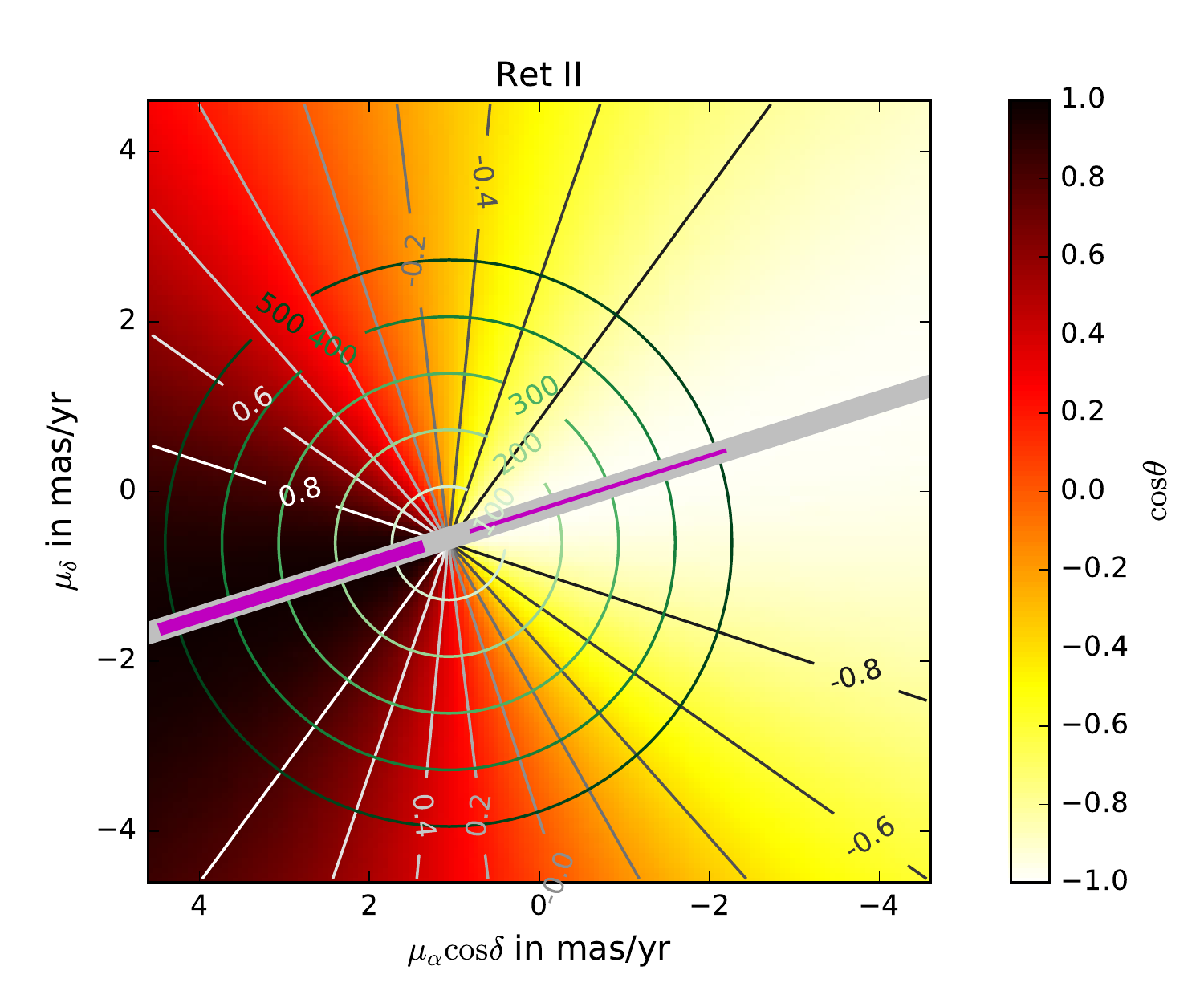}
 \includegraphics[width=0.32\textwidth,clip]{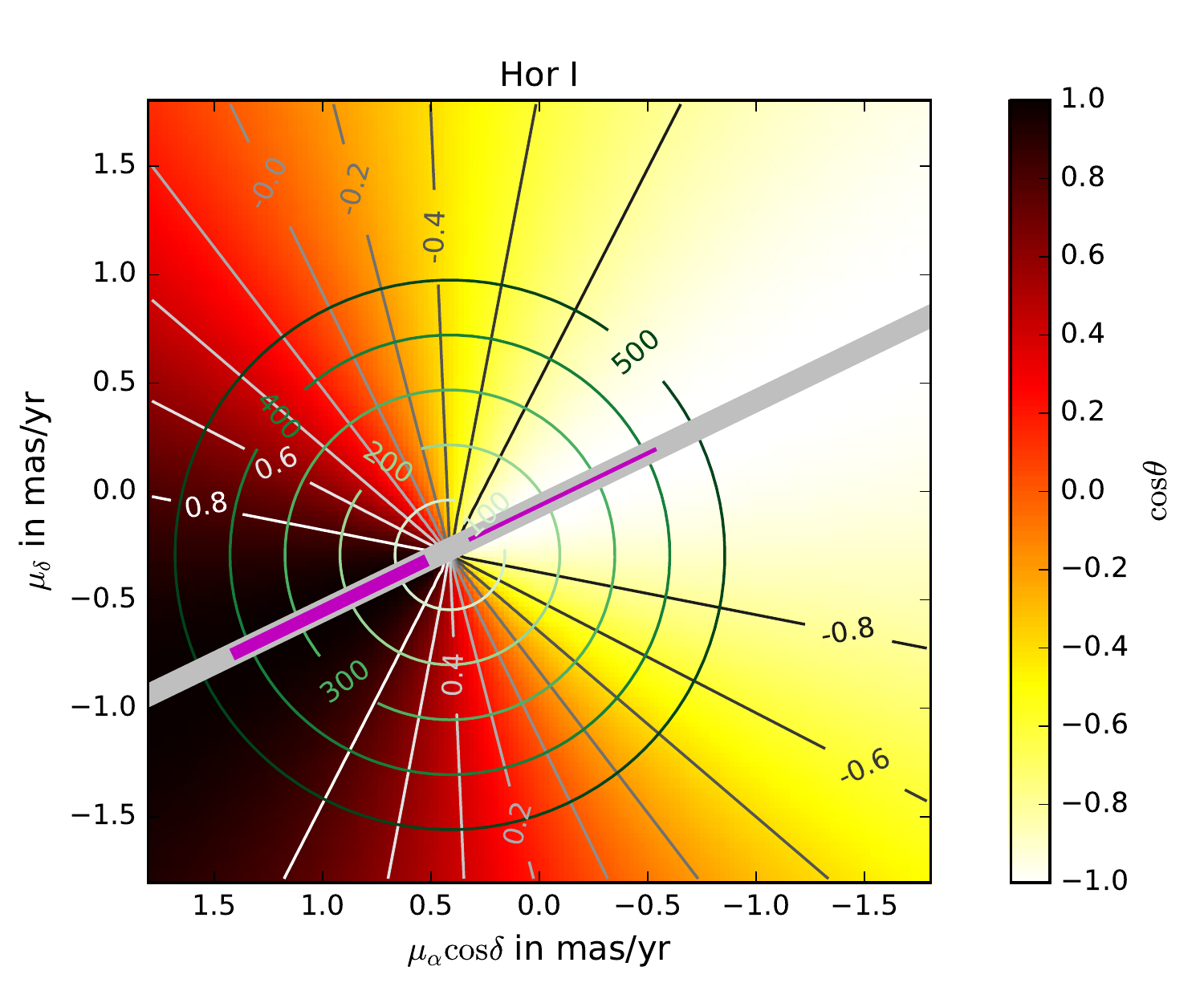}      
 \includegraphics[width=0.32\textwidth,clip]{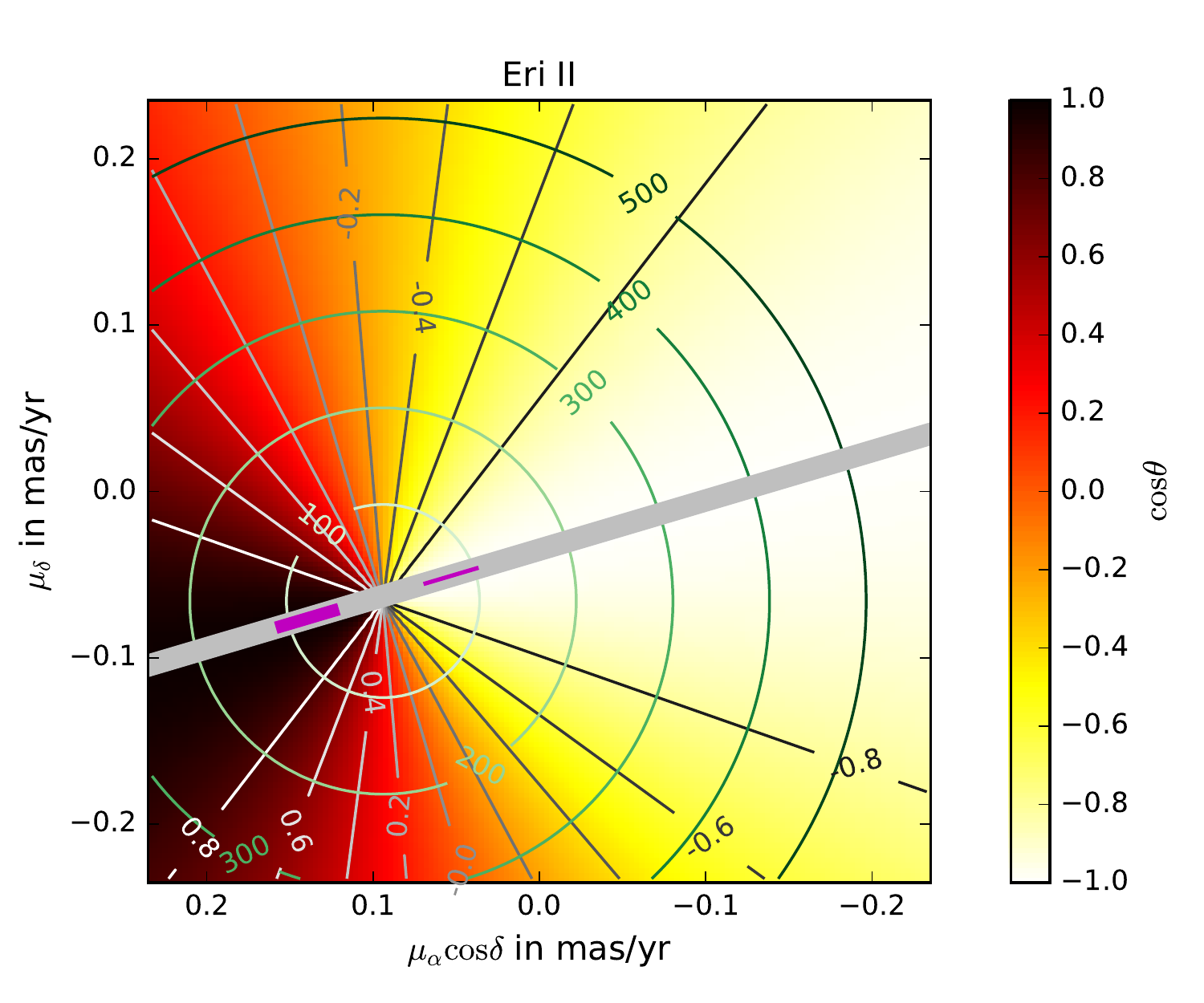}      
%% Note the ABSENCE of the extension .pdf  !
  \caption{Predicted proper motions and expected Gaia proper motion accuracy (cyan circle) for a single star at the detection limit of $V = 20\,\mathrm{mag}$, for three faint MW satellite galaxies discovered in the DES survey \citep{Koposov2015}: {\bf Left:} Reticulum\,II at 30\,kpc distance, {\bf Middle:} Horologium\,I at 80\,kpc distance, {\bf Right:} Eridanus\,II at 380\,kpc distance. Plotted are the two proper motion coordinates, with the angle between the resulting orbital plane and the VPOS plane color-coded and indicated by radial grey contours. The circular green contours indicate the resulting absolute speed for each possible proper motion combination. The predicted proper motions (resulting in the best alignment of the orbit with the VPOS) are indicated as a thick magenta line (co-orbiting with the majority of classical satellites) and thin magenta line (counter-orbiting). }
  \label{pawlowski2:fig3}
\end{figure}

The predicted proper motions range from $\approx 1 \mathrm{mas\,yr}^{-1}$\ to less than  $\approx 100 \mu\mathrm{as\,yr}^{-1}$, depending on the distance of the satellite. Such proper motion accuracy is achievable with Hubble Space Telescope (HST) measurements, and possibly also by ground-based observations given sufficiently long temporal baselines \citep{Casetti-Dinescu2017}. This is in part due to large-number statistics in the stars and background reference sources. Current measurements, several of which became available only after the satellite proper motions motion were predicted in \citet{Pawlowski2013}, indicate that these predictions are well met by the 11 classical satellite galaxies. In particular, more accurate measurement tend to result in a better agreement with the prediction (see Figure \ref{pawlowski2:fig2}).

Figure \ref{pawlowski2:fig3} shows predicted proper motions for three examples of MW satellites. These three satellites are part of those recently discovered by \citet{Koposov2015} using data from the Dark Energy Survey (DES), and lie at different distances. Checking the Color-Magnitude Diagrams of the faint satellites in \citet{Koposov2015} shows that the more nearby satellites do potentially have a few ($\leq 10$) stars with apparent magnitudes above the Gaia limit of  $V = 20\,\mathrm{mag}$. It might thus be possible to obtain systemic proper motions for these systems with Gaia from a few individual stars, instead of a large sample of stars from deeper HST observations and long temporal baselines. Would Gaia provide sufficient information to test the proper motion predictions and thus help us learn more about the VPOS?

As an estimate for the Gaia proper motion accuracy we use the post-launch predictions by \citet{deBruijne2014}: $\Delta\mu \approx 330 \mu\mathrm{as\,yr^{-1}}$\ as expected for a single star with $V = 20\,\mathrm{mag}$\ and a color equivalent to a G2V star. At this accuracy, internal motions in the satellite galaxies can be neglected, in particular for the faintest satellites which have velocity dispersions of $\leq 10\,\mathrm{km\,s^{-1}}$, which is an order of magnitude smaller than their orbital velocities.
A comparison of this estimated accuracy (cyan circles) with the predicted proper motions in Figure \ref{pawlowski2:fig3} illustrates that Gaia can provide good constraints on the most nearby satellite galaxies individually (left panel). For satellites at intermediate distances, Gaia proper motions will not constrain the orbits of individual satellites well. However, Gaia can provide proper motions for many such satellites. Such a larger sample of satellite galaxies with proper motions, albeit less certain, will allows to statistically test whether there are indications for a preferred kinematic correlation with the VPOS. For very distant satellites, Gaia will clearly not provide any proper motion constraints, both due to the small expected proper motions and the large distance that results in even the brightest stars in the satellites being too faint for Gaia.

In addition to testing the kinematic coherence of the VPOS via proper motions, Gaia also holds the potential to uncover stellar streams, which can be compared to the orientation of the VPOS and thus test if the preferential stream alignment persists. As discussed in \citet{Pawlowski2012}, this will have to take into account the preference to detect nearby streams: since the VPOS is almost perpendicular to the Sun's position relative to the Galactic center, only streams at distances beyond $\approx 8\,\mathrm{kpc}$\ can possibly align with it. Our position thus biases away from finding aligned streams, which makes the already known association of up to half of the known MW streams all the more compelling.

\section{Conclusions}
%%--------------------

Known satellite galaxies, globular clusters, and streams around the MW preferentially align in a flattened distribution, the Vast Polar Structure (VPOS). The preferential alignment of streams in the MW halo indicates that objects preferentially move along this structure. This is confirmed by orbits inferred from proper motion measurements for the 11 classical satellite galaxies, most of which co-orbit along the VPOS. The existence of the VPOS thus allows to empirically predict the likely range of proper motions for most MW satellites, based on the assumption that their orbits, too, align with the structure. Testing these predictions for a sample of satellite galaxies thus allows to determine how strongly rotationally supported the VPOS is, and whether there are differences in the orbital alignment between the brightest and fainter MW satellites. While individual HST proper motion measurements can deliver accurate systemic proper motions for distant satellite galaxies, Gaia provides a complementary approach. Its extensive sky coverage and accuracy can provide proper motion constraints even for faint satellite galaxies, if the few brightest stars belonging to these objects can be accurately identified. While the accuracy for single-star proper motions will not be sufficient for precise orbit determinations for individual satellites, Gaia will allow to test how closely the full sample of satellite galaxies follows the predicted proper motions based on the assumption of orbits aligned with the VPOS.

% Optional acknowledgements
% -------------------------
\begin{acknowledgements}
Support for this work was provided by NASA through Hubble Fellowship grant \#HST-HF2-51379.001-A awarded by the Space Telescope Science Institute, which is operated by the Association of Universities  for  Research  in  Astronomy,  Inc.,  for  NASA,  under  contract  NAS5-26555
\end{acknowledgements}

%%-----------------------------
%%   Bibliography
%%-----------------------------
%%

%% The following lines are required when using BibTEX (strongly encouraged!):
\bibliographystyle{aa}  % A&A bibliography style file (aa.bst)
\bibliography{pawlowski2} % your references in file: Yourfile.bib

\begin{thebibliography}{12}
\expandafter\ifx\csname natexlab\endcsname\relax\def\natexlab#1{#1}\fi

\bibitem[{{Casetti-Dinescu} {et~al.}(2017){Casetti-Dinescu}, {Girard}, \&
  {Schriefer}}]{Casetti-Dinescu2017}
{Casetti-Dinescu}, D.~I., {Girard}, T.~M., \& {Schriefer}, M. 2017, ArXiv
  e-prints

\bibitem[{{de Bruijne} {et~al.}(2014){de Bruijne}, {Rygl}, \&
  {Antoja}}]{deBruijne2014}
{de Bruijne}, J.~H.~J., {Rygl}, K.~L.~J., \& {Antoja}, T. 2014, in EAS
  Publications Series, Vol.~67, EAS Publications Series, 23--29

\bibitem[{{Grillmair}(2017)}]{Grillmair2017}
{Grillmair}, C.~J. 2017, \apj, 847, 119

\bibitem[{{Koposov} {et~al.}(2015){Koposov}, {Belokurov}, {Torrealba}, \&
  {Evans}}]{Koposov2015}
{Koposov}, S.~E., {Belokurov}, V., {Torrealba}, G., \& {Evans}, N.~W. 2015,
  \apj, 805, 130

\bibitem[{{Kunkel} \& {Demers}(1976)}]{Kunkel1976}
{Kunkel}, W.~E. \& {Demers}, S. 1976, in Royal Greenwich Observatory Bulletins,
  Vol. 182, The Galaxy and the Local Group, ed. R.~J. {Dickens}, J.~E. {Perry},
  F.~G. {Smith}, \& I.~R. {King}, 241

\bibitem[{{Lynden-Bell}(1976)}]{Lynden-Bell1976}
{Lynden-Bell}, D. 1976, \mnras, 174, 695

\bibitem[{{Metz} {et~al.}(2008){Metz}, {Kroupa}, \& {Libeskind}}]{Metz2008}
{Metz}, M., {Kroupa}, P., \& {Libeskind}, N.~I. 2008, \apj, 680, 287

\bibitem[{{Pawlowski} \& {Kroupa}(2013)}]{Pawlowski2013}
{Pawlowski}, M.~S. \& {Kroupa}, P. 2013, \mnras, 435, 2116

\bibitem[{{Pawlowski} \& {Kroupa}(2014)}]{PawlowskiKroupa2014}
{Pawlowski}, M.~S. \& {Kroupa}, P. 2014, \apj, 790, 74

\bibitem[{{Pawlowski} {et~al.}(2015){Pawlowski}, {McGaugh}, \&
  {Jerjen}}]{Pawlowski2015}
{Pawlowski}, M.~S., {McGaugh}, S.~S., \& {Jerjen}, H. 2015, \mnras, 453, 1047

\bibitem[{{Pawlowski} {et~al.}(2012){Pawlowski}, {Pflamm-Altenburg}, \&
  {Kroupa}}]{Pawlowski2012}
{Pawlowski}, M.~S., {Pflamm-Altenburg}, J., \& {Kroupa}, P. 2012, \mnras, 423,
  1109

\bibitem[{{Sohn} {et~al.}(2017){Sohn}, {Patel}, {Besla}, {van der Marel},
  {Bullock}, {Strigari}, {van de Ven}, \& {Walker}}]{Sohn2017}
{Sohn}, S.~T., {Patel}, E., {Besla}, G., {et~al.} 2017, ArXiv e-prints

\end{thebibliography}

\end{document}